\documentclass[iop,apj,twocolumn,numberedappendix,appendixfloats]{openjournal}
\usepackage{fix-cm}
\usepackage{lastpage}
\usepackage{graphicx}
\graphicspath{ {./figures/} }
\usepackage{xcolor}

\defcitealias{oehl2025a}{OT25}
\usepackage{newtxtext,newtxmath,enumitem,multirow,ctable,nccmath,tensor,amsmath,booktabs}
\usepackage{bm}

\usepackage{bbold}
\usepackage{savesym}
\savesymbol{tablenum}
\usepackage{siunitx}

\usepackage{hyperref}
\hypersetup{breaklinks,colorlinks,citecolor=blue,urlcolor=cyan}
\usepackage[capitalise]{cleveref}
\crefname{section}{Sec.}{Secs.}
\crefname{subsection}{Sec.}{Secs.}

\usepackage{blkarray}
\newcommand{\dd}{\mathrm{d}}
\newcommand{\bvec}[1]{\boldsymbol{\mathbf{#1}}}
\renewcommand{\pi}{\uppi}
\newcommand{\omegam}{\Omega_{\mathrm{m}}}

\DeclareSIUnit\parsec{pc}
\DeclareSIUnit\sqd{deg^2}
\DeclareSIUnit\arcmin{arcmin}

\shorttitle{Non-Gaussian weak lensing likelihood}
\shortauthors{V. Oehl \& T. Tröster}
\begin{document}

\title{The Non-Gaussian Weak-Lensing Likelihood: A Multivariate Copula Construction and Impact on Cosmological Constraints}
\author{Veronika Oehl$^{\star}$}
\author{Tilman Tröster}
\affiliation{Institute for Particle Physics and Astrophysics, ETH Zurich, 8093 Zurich, Switzerland}
\thanks{$^\star$E-mail: \href{mailto:veoehl@phys.ethz.ch}{veoehl@phys.ethz.ch}}

\journalinfo{The Open Journal of Astrophysics}
\submitted{Accepted 2026 June 11}

\begin{abstract}
We present a framework to compute non-Gaussian likelihoods for two-point correlation functions.
The non-Gaussianity is most pronounced on large scales that will be well-measured by stage-IV weak-lensing surveys.
We show how such a multivariate likelihood can be constructed and efficiently evaluated using a copula approach by incorporating exact one-dimensional marginals and a dependence structure derived from the exact multivariate likelihood.
The copula likelihood is found to be in better agreement with simulated sampling distributions of correlation functions than Gaussian likelihoods, particularly on large scales.
We furthermore investigate the effect of the non-Gaussian copula likelihood on posterior inference, including sampling the full parameter space of contemporary weak-lensing analyses.
We find potential parameter shifts in $S_8$ on the order of one standard deviation for \SI{1000}{\sqd} surveys but negligible shifts for areas of \SI{10000}{\sqd}, suggesting Gaussian likelihoods are sufficient for stage-IV surveys, though results depend on the detailed mask geometry and data vector structure.

\end{abstract}
\keywords{likelihood, non Gaussian, Bayesian statistics, Gaussian random field, correlation function, weak lensing}

\section{Introduction}
Current and upcoming cosmological surveys aim for sub-percent-level parameter constraints. 
With this increase in precision comes the need for revisiting assumptions in the statistical modeling.
Accurate modeling of observables requires not only precise theoretical predictions but also a faithful description of their associated uncertainties, encoded in the likelihood for a given parametric model.
In most applications, the likelihood is assumed to be Gaussian, which is often a good approximation.
It has been shown however that this approximation breaks down in certain regimes, for example for $n$-point functions of Gaussian random fields, where the validity of the Gaussian likelihood assumption depends on survey conditions and the angular scales considered \citep[e.g.][]{amendola1996a,hamimeche2008,sellentin2017,sellentin2018,lin2020a,upham2020,hall2022a,oehl2025a}.

Real-space correlation functions are widely used two-point statistics for cosmic shear.
Their likelihood is approximated as Gaussian in recent surveys \citep{asgari2021a,li2023a,wright2025a,collaboration2026a}.
The central limit theorem is often invoked to justify this, but this assumption may not hold in practice: the sampling distributions of weak-lensing correlation functions have been shown to be non-Gaussian, especially on large scales \citep{sellentin2017,sellentin2018,joachimi2021a} but explicit frameworks to compute a more exact likelihood are still missing.

In this work, we will show how to compute an explicit non-Gaussian likelihood for weak-lensing correlation function data vectors on arbitrarily weighted Gaussian random fields.
There have been attempts to construct non-Gaussian likelihoods, particularly for weak lensing power spectra \citep{sato2011a,upham2020,lin2020a}, but so far no parameter constraints from multidimensional correlation-function data vectors have been obtained from such non-Gaussian likelihoods.

A straightforward way to introduce non-Gaussianity to the likelihood is by taking into account higher order moments in its Edgeworth expansion \citep{amendola1996a,keitel2011,collaboration2026b}.
This approach does however have fundamental limitations: it can yield negative densities and is not guaranteed to be a proper probability distribution.
What is more, the expansion is only accurate in the vicinity of the expansion point, and adding more terms does not necessarily improve the approximation globally.

For our framework, we leverage exact one-dimensional marginals for the weak-lensing correlation function on masked skies from \citet[hereafter OT25]{oehl2025a}.
We then use a copula to lift these marginals to a multivariate likelihood.
The exact framework from \citetalias{oehl2025a} technically does allow setting up full multivariate distributions but it is computationally infeasible to evaluate these likelihoods beyond low-dimensional examples.
Copulas provide a flexible, mathematically well-founded way to construct multivariate distributions from known marginals and a dependence structure \citep{nelsen1999a,joe2014a}.
They allow us to use our exact non-Gaussian marginals, while modeling dependence between correlation-function data points tractably.

Copula models have been widely used in different fields such as survival analysis \citep{oakes1989a}, econometrics \citep{trivedl2007a}, climate research or finance \citep{genest2009a}.
A lot of these works explore how to estimate the copula from data \citep{min2010a,smith2013a}, while there are only few examples applying copulas with known analytic non-Gaussian marginals as likelihoods.
In cosmology, a few works have suggested using copulas.
Early examples include \citet{scherrer2009a}, who studied the copula of the evolved dark-matter density field.
Both \citet{benabed2009a} and \citet{sato2010a,sato2011a} use copulas as likelihoods for two-point functions but both are for power spectra and in both cases the marginal distributions are fit to simulations. 
\citet{wilking2013a} explore the copula with analytic marginals from \citet{keitel2011} as alternative approach to their work. 
Both works are on one dimensional fields however and some copula parameters are estimated from simulations.
Recently, \citet{uhlemann2023a} suggested using copulas as a potential method to construct the joint probability density function of convergence one-point functions.

For the first time, we apply a copula likelihood with analytical non-Gaussian marginals to a full weak-lensing correlation-function data vector.
Based on our previous work, we show how to compute the exact marginals, the covariance and how they are combined using a Gaussian copula.
We systematically compare the resulting posteriors with those from a standard Gaussian likelihood in a weak-lensing case study using mock data vectors constructed for a setup similar to the Kilo-Degree Survey \citep[KiDS-1000,][]{asgari2021a} employing survey masks approximating the footprint of stage-III surveys such as the Subaru Hyper Suprime-Cam \citep[HSC,][]{aihara2018a,dalal2023a,li2023a} survey, the Dark Energy Survey \citep[DES,][]{collaboration:2016a,secco2022a,amon2022a} or KiDS-1000 \citep{kuijken2019a,asgari2021a,li2023b} as well as stage-IV surveys such as the Vera C. Rubin Observatory Legacy Survey of Space and Time \citep[LSST,][]{collaboration2009a}, Euclid \citep{laureijs2011a} or the Nancy Grace Roman Space Telescope \citep{spergel2015a}.

We show that the likelihood choice can materially influence the resulting posteriors and shift their means by a significant fraction of one standard deviation compared to the Gaussian case.
This paper is structured as follows:
In Sec.~\ref{sec:theory}, we summarize the theory behind the exact correlation-function likelihood on masked Gaussian random fields. 
Sec.~\ref{sec:copula} introduces the copula by the example of a Gaussian dependence structure as we use it for most of this work.
Implementation details are shown in Sec.~\ref{sec:implementation} together with some approximations that need to be made.
In Sec.~\ref{sec:application}, we apply our framework to a realistic weak-lensing setup, compare the copula probability distribution to sampling distributions and compute posteriors on cosmological parameters for different survey masks.
We close with a discussion of the results and a conclusion.

\section{The exact correlation function likelihood}\label{sec:theory}
In this section, we recap the main results from \citetalias{oehl2025a} that are necessary to set up the copula likelihood for weak-lensing correlation functions.
As the covariance structure of a Gaussian random field is directly given by the power spectrum in harmonic space, the likelihood of the real-space correlation function is more easily worked out in terms of the spherical harmonic coefficients $a_{\ell m}$.
For a masked field, i.e. when parts of the field are set to zero, the resulting coefficients are usually referred to as pseudo-$a_{\ell m}$, which we will denote with a tilde in equations.
We start by writing the correlation-function estimator as a quadratic form in the pseudo-$a_{\ell m}$ \citepalias[see][for details]{oehl2025a}
\begin{equation}
    \label{eq:xip_alm}
        \hat{\xi}^{+}_{ij} \left(\bar{\theta}\right) =  \sum_{\ell} K_{\ell} \left(\bar{\theta}\right)  \sum_{m=-\ell}^{\ell} \left(\tilde{a}^{E,i}_{\ell m} {\tilde{a}^{\ast \ E,j}_{\ell m}} +  \tilde{a}^{B,i}_{\ell m} {\tilde{a}^{\ast \ B,j}_{\ell m}}\right),
\end{equation}
where $i$ and $j$ label redshift bins, and $E$ and $B$ denote $E$- and $B$-modes of the spin-2 weak lensing field.
The angular dependence of the correlation function is completely contained in $K_{\ell} \left(\bar{\theta}\right)$ as
\begin{equation}
\label{eq:k_ell}
    K_{\ell} \left(\bar{\theta}\right) = 2 \pi \  \ \frac{2}{\theta_{\mathrm{max}}^2 - \theta_{\mathrm{min}}^2} \int_{\theta_{\mathrm{min}}}^{\theta_{\mathrm{max}}} \dd \theta \ \theta B(\theta) d_{2 2}^{\ell} (\theta),
\end{equation}
where $\bar{\theta}$ is an angular separation bin with bin edges $[\theta_{\mathrm{min}}, \theta_{\mathrm{max}}]$ and $B(\theta) = \left(2 \pi \sum_{\ell} \left(2 \ell + 1 \right) P_{\ell} \left(\cos \theta \right) w_{\ell}\right)^{-1}$ with $w_{\ell}$, the mask power spectrum and $P_{\ell}$ the Legendre polynomials.
Our correlation-function estimator, \cref{eq:xip_alm}, can compactly be written as 
\begin{equation}
    \label{eq:xip_M_bin}
    \hat{\xi}^{+}_{ij} (\bar{\theta}) = \tilde{\bvec{a}}^T \bvec{M}_{ij}^{\xi^{+}}(\bar{\theta}) \tilde{\bvec{a}},
\end{equation}
where we call $\bvec{M}_{ij}^{\xi^{+}}(\bar{\theta})$ the combination matrix for $\xi^+$ for a given redshift-bin combination $ij$ and an angular-separation bin $\bar{\theta}$.
Bold $\tilde{\bvec{a}}$ are the stacked pseudo-$a_{\ell m}$.
As established in \citetalias{oehl2025a} and \citet{upham2020} before, the pseudo-$a_{\ell m}$ of an underlying Gaussian random field are Gaussian distributed.
The pseudo-$a_{\ell m}$ covariance $\bvec{\Sigma}$ can be derived using the full-field covariances given by the power spectra $C_{\ell}$, which depend on the cosmological parameters, and the mask properties \citepalias[see][equation (24)]{oehl2025a}.

For a dataset of correlation-function values $\hat{\xi}^{+}_{ij} (\bar{\theta}_k)$ we introduce a mapping $\{ijk\} \rightarrow q$, so each combination of redshift bins and angular separation corresponds to one data dimension and $q$ runs from $1$ to $n_d$, the total number of dimensions of the data vector.
An $n_d$-dimensional likelihood for a set of quadratic forms $\tilde{\bvec{a}}^T \bvec{M}_q^{\xi^{+}} \tilde{\bvec{a}}$ in Gaussian random variables $\tilde{\bvec{a}}$ with covariance $\bvec{\Sigma}$ is given by the inverse Fourier transform of the characteristic function
\begin{equation}
\label{eq:char_num}
    \varphi_{\xi^+} (\bvec{t}) = \prod_j (1-2i\lambda_j)^{-1/2},
\end{equation}
where the $\lambda_j$ are eigenvalues of the sum over products of combination matrices $\bvec{M}^{\xi^+}_q$ and the pseudo-$a_{\ell m}$ covariance matrix $\bf{\Sigma}$ for each dimension $q$
\begin{equation}
\label{eq:eigvals}
    \lambda_j \in \lambda \left(\sum_q t_q \bvec{M}^{\xi^+}_q \bf{\Sigma}\right)
\end{equation}
and $\bvec{t}$ is an $n_d$-dimensional grid in the characteristic function space.
We refer to \citetalias{oehl2025a} for details.
For future reference, we derive the $n_d$-dimensional mean and covariance in the remainder of this section.
For this use case, it is convenient to write the characteristic function equivalently as 
\begin{equation}
\label{eq:char_ana}
    \varphi_{\xi^+} (\bvec{t}) = \vert \mathbb{1} - 2i \sum_q t_q \bvec{M}^{\xi^+}_q \bvec{\Sigma} \vert^{-1/2},
\end{equation}
see \citet{upham2020} for a proof.

The moments of any distribution can be computed by taking derivatives of the characteristic function with respect to the characteristic function parameter $t$:
\begin{equation}
\label{eq:moments}
    \varphi_{\xi^+}^{(j)}(t=0) = i^j \operatorname{E}[\xi^j],
\end{equation}
where $^{(j)}$ denotes the $j$-th derivative with respect to $t$. 

\subsection{Mean}
\label{sec:mean}
Applying \cref{eq:moments} to \cref{eq:char_ana}, the means for the correlation function likelihood can be calculated as
\begin{equation}
\label{eq:mu}
    \mu\left(\xi^+_q\right) = \operatorname{tr} \left(\bvec{M}^{\xi^+}_q \bvec{\Sigma}\right).
\end{equation}
The mean can also be calculated by using \cref{eq:xip_alm} directly, but we will show in Sec.~\ref{sec:implementation} why the representation in \cref{eq:mu} is useful. 
We have shown in \citetalias{oehl2025a} that \cref{eq:mu} correctly recovers \cref{eq:xip_alm}.

\subsection{Covariance}
\label{sec:cov}
For the correlation-function covariance, one requires second derivatives and the basic formula for the moments can be extended to a multi-dimensional version using partial derivatives
\begin{equation}
    \operatorname{E}[\xi^+_i \xi^+_j] = -  \frac{\partial \varphi_{\xi^+}}{\partial t_i \partial t_j} (\bvec{t} = \bvec{0}).
\end{equation}
Defining $\mathbb{1} - 2i \sum_q t_q \bvec{M}_q \bvec{\Sigma} \equiv \bvec{A}$, we will make use of Jacobi's formula
\begin{equation}
    \frac{d}{dt} \det \bvec{A}(t) = \left(\det \bvec{A}(t) \right) \cdot \operatorname{tr} \left (\bvec{A}(t)^{-1} \cdot \, \frac{\mathrm{d}\bvec{A}(t)}{\mathrm{d}t}\right ).
\end{equation}
We note that
\begin{equation}
    \frac{\partial \bvec{A}}{\partial t_j} \biggr\rvert_{\bvec{t} = \bvec{0}} = -2 i \bvec{M_j} \bvec{\Sigma}.
\end{equation}
This yields
\begin{multline}
    \frac{\partial \varphi_{\xi^+}}{\partial t_i \partial t_j} = \vert \bvec{A} \vert^{-1/2} \left(\frac{1}{4} \operatorname{tr}\left(\bvec{A}^{-1} \frac{\partial \bvec{A}}{\partial t_i}  \right) \operatorname{tr}\left(\bvec{A}^{-1} \frac{\partial \bvec{A}}{\partial t_j}\right) \right.\\  \left. - \frac{1}{2} \operatorname{tr} \left( -\bvec{A}^{-1} \frac{\partial \bvec{A}}{\partial t_j} \bvec{A}^{-1} \frac{\partial \bvec{A}}{\partial t_i}  \right)\right).
\end{multline}
    
As $\bvec{t}$ only appears in the sum in \cref{eq:char_ana}, taking derivatives separates the dimensions to our advantage.
Noting that $\bvec{A}^{-1} (\bvec{t} = \bvec{0}) = \mathbb{1}$ and $\vert \bvec{A} \vert (\bvec{t} = \bvec{0}) = 1$\footnote{Note that all operations like matrix multiplication, taking the determinant and inverse commute with setting $\bvec{t} = 0$. For taking the inverse this holds if $\bvec{A} = \bvec{A}_0 + t \bvec{A}_1$, where $\bvec{A}_0$ must be invertible. This is the case for our definition of $\bvec{A}$.} one can compute 
\begin{equation}
\label{eq:xicov}
    \operatorname{E}[\xi_i^+ \xi_j^+] = \operatorname{tr} \left( \bvec{M}_i^{\xi^+} \bvec{\Sigma} \right) \operatorname{tr} \left( \bvec{M}_j^{\xi^+} \bvec{\Sigma} \right) + 2 \operatorname{tr} \left( \bvec{M}_i^{\xi^+} \bvec{\Sigma} \bvec{M}_j^{\xi^+} \bvec{\Sigma}\right).
\end{equation}
This is the general, non-central second moment.
Subtracting the product of the means, we arrive at the covariance of correlation functions
\begin{equation}
\label{eq:xicov}
    \bvec{\Sigma}_{\xi} = \operatorname{cov} \left(\xi_i^+, \xi_j^+\right) = 2 \operatorname{tr} \left( \bvec{M}_i^{\xi^+} \bvec{\Sigma} \bvec{M}_j^{\xi^+} \bvec{\Sigma}\right).
\end{equation}
A similar result was derived in \citet{dahlen2008a}.
Naturally, this covariance inherits a parameter dependence through the pseudo-$a_{\ell m}$ covariance $\bvec\Sigma$. 
In the following, we retain this parameter dependence for the exact likelihood construction, while for the Gaussian likelihood comparison we use a covariance fixed at the fiducial cosmology.
The sparsity of the combination matrices $\bvec{M}$ and the trace operator will allow for substantial numerical simplifications which we will come back to in Sec.~\ref{sec:implementation}

\section{The copula approach}
\label{sec:copula}
Building $n_d$-dimensional likelihoods using the full characteristic function is computationally infeasible as eigenvalues need to be retrieved $N^{n_d}$ times, where $N$ is the number of points in the characteristic function space in each dimension.
This is not only infeasible but also very inefficient for inferences because the whole probability density function is calculated explicitly even if only a point-evaluation of the likelihood would be needed. 
Copulas are constructed from one-dimensional marginal distributions and a coupling density.
The copula therefore reduces the coupling between dimensions to the coupling of a simpler distribution, only employing the covariance, while the exact one-dimensional marginals are recovered. 
This makes the $n_d$-dimensional calculation, which is the crucial scaling escalating the computational requirements, unnecessary and also allows for point-wise evaluation.

\subsection{Copula theory}
Previous work \citep{sato2011a} introduced copula-based likelihood approaches in weak lensing.
In that context, $\chi^2$ distributions with their mean and standard deviation estimated from simulations were used as marginals to describe the likelihood of power spectra measured on a \SI{25}{\sqd} sky patch. 
However, this distribution is only accurate on the full sky, such that the fit of the $\chi^2$-marginals is an approximation to the true distribution.
Here, we leverage our ability to compute exact one-dimensional marginal likelihoods $p(\xi^+_q)$ for the correlation function on an arbitrarily masked sky in each data dimension $q$.
Each marginal likelihood $p(\xi^+_q)$ can be evaluated at the corresponding part of the data vector $\xi^+_q$, $p(\xi^+_q \vert \vartheta)$, where $\vartheta$ is a set of model parameters.
Conceptually, a copula works by mapping each one-dimensional marginal $p(\xi^+_q)$ to a simpler reference distribution using its cumulative distribution function.
One then defines an $n_d$-dimensional coupling distribution of the same type, using the covariance structure of the exact distribution.
This would be $\bvec{\Sigma}_{\xi}$, \cref{eq:xicov}, for our correlation-function case.

We illustrate the construction of a copula using a Gaussian dependence model, which we also adopt for the analysis in this work.
For a more general introduction to the topic see for example \citet[][]{nelsen1999a}.
We define the vector
\begin{equation}
\label{eq:normspacevec}
   z_q = c^{-1}_{\mathrm{norm}}\left(c (\xi^+_q)\right),
\end{equation}
where $c^{-1}_{\mathrm{norm}}$ is the inverse of the normal cumulative distribution function (CDF), also known as percent point function (PPF) and $c (\xi^+_q)$ is the CDF of $p (\xi^+_q)$. 

The CDFs $c(\bvec{\xi^+})$ can directly be evaluated at the data vector $\bvec{\xi^+}$ such that $c^{-1}_{\mathrm{norm}}$ only needs to be computed at $n_d$ points.
The $c (\xi^+_q)$ are computed by classical numerical trapezoidal integration and then interpolated to evaluate at a data point $\xi^+_q$.

The dependence structure is modelled using a copula density.
The construction starts from a multivariate normal distribution with covariance matrix $\bvec{C}_{\xi}$, which is the correlation matrix corresponding to $\bvec{\Sigma}_{\xi}$, \cref{eq:xicov}, evaluated at the vector $\bvec{z}$, \cref{eq:normspacevec}. In this example, this is a multivariate normal distribution
\begin{equation}
    p_{\mathrm{MV}} = \mathcal{N} \left(\bm{z}, \, \bvec{C}_{\xi}\right).
\end{equation}
This multivariate probability density needs to be normalized by the product over one-dimensional standard normal PDFs evaluated at the same points $\bvec{z}$ to yield the overall Gaussian copula density:
\begin{equation}
    \rho_{\mathrm{Gauss}} = p_{\mathrm{MV}} / \prod_q p_{\mathrm{norm}} (z_q).
\end{equation}
Intuitively, the copula density equals one when the dimensions are independent, so the multivariate likelihood simply reduces to the product of the one-dimensional likelihoods.
Multiplying the copula density with the one-dimensional non-Gaussian marginals evaluated at each data point transforms them into the full multivariate non-Gaussian likelihood:
\begin{equation}
    p(\bvec{\xi^+} \vert \vartheta) = \rho_{\mathrm{Gauss}} \ \prod_q p(\xi^+_q \vert \vartheta). 
\end{equation}
Alternative coupling structures, such as stronger tail dependence, can be modeled by replacing the Gaussian copula density with another copula. The procedure described above remains unchanged, except that the inverse CDF and the corresponding multi- and univariate distributions of the chosen copula are used instead of the normal distributions.
The difficulty of using other distributions lies in the fact that many of the well-known distributions do not have a well-defined multivariate extension, however.
Apart from that, choosing or fitting hyperparameters to fix additional characteristics of other coupling distributions might be necessary, as in the case of a Student-t distribution, where the degrees of freedom need to be specified.

\section{Implementation}
\label{sec:implementation}

In this section, we detail key aspects of the implementation of the framework for its application to a weak-lensing correlation-function data vector.
The copula likelihood framework is implemented in the publicly available Python package \texttt{xilikelihood}\footnote{\url{https://github.com/voehl12/xilikelihood}}.
As explained in Sec.~\ref{sec:copula}, the copula construction requires the one-dimensional marginals $p(\bvec{\xi^+})$ and the covariance matrix $\bvec{\Sigma}_{\xi}$. 
For both, the structures of the combination matrix $\bvec{M}_q^{\xi^+}$ and of the pseudo-$a_{\ell m}$ covariance matrix $\bvec{\Sigma}$ play an important role.  
The pseudo-$a_{\ell m}$ covariance matrix $\bvec{\Sigma}$ for a full dataset of correlation functions is a block matrix, where each block describes the covariance of two sets of pseudo-$a_{\ell m}$. 
These two sets describe the fields generated by two different source redshift-distributions $z_i$ and $z_j$. 
We therefore denote each block as $\bvec{B}_{z_i,z_j}$ and write the pseudo-$a_{\ell m}$ covariance matrix as:
\begin{equation}
    \label{eq:palm_cov}
\bvec{\Sigma}  =
    \begin{pmatrix}
        \bvec{B}_{z_1,z_1}& \bvec{B}_{z_1,z_2} & \dots & & \dots & \bvec{B}_{z_1,z_n}\\
        \bvec{B}_{z_2,z_1} & \ddots &  & & & \vdots \\
        \vdots &  & \bvec{B}_{z_i,z_i} &  & & \vdots \\
        \vdots &  & & & \ddots & \bvec{B}_{z_{n-1},z_n}\\
        \bvec{B}_{z_n,z_1} & \dots & & \dots & \bvec{B}_{z_n,z_{n-1}} & \bvec{B}_{z_n,z_n}.
    \end{pmatrix}
\end{equation}
The combination matrices $\bvec{M}_q^{\xi^+}$ select and sum over the pseudo-$a_{\ell m}$ needed for a given redshift-bin combination.
They are therefore sparse and block-diagonal.
We write a combination matrix for a cross correlation between redshift bins $i$ and $j$ and angular-separation bin $\theta_k$, corresponding to data point $q$ as
\begin{equation}
    \label{eq:comb_mat}
 \bvec{M}_{q}^{\xi^+}  =
    \begin{blockarray}{cccccc}
   & \cdots & i & \cdots & j & \cdots \\
\begin{block}{c(ccccc)}
\vdots & 0 & \cdots & 0 & \cdots & 0 \\
i      & 0 & \cdots &  & \bvec{m}_{\theta_k} & 0 \\
\vdots & \vdots & \ddots & \vdots & & \vdots \\
j      & 0 & \bvec{m}_{\theta_k} &  & \cdots & 0 \\
\vdots & 0 & \cdots & 0 & \cdots & 0 \\
\end{block}
\end{blockarray}.
\end{equation}
Here, the $\bvec{m}_{\theta}$ denote diagonal matrices whose entries run over all multipoles $\ell$ and modes $m$, containing the $\ell$-dependent prefactor in \cref{eq:xip_alm}.
Each product $\bvec{M}_q^{\xi^+} \bvec{\Sigma}$ is consequently a series of row-wise multiplications of a combination vector $\bvec{m}_{\theta_k}$ with a sub-block $\bvec{B}_{z_i,z_j}$ of $\bvec{\Sigma}$.

\subsection{Marginals}
The marginals are obtained from the characteristic function, \cref{eq:char_num}.
Because only one-dimensional characteristic functions are evaluated, the sum in \cref{eq:eigvals} reduces to a single term, and the grid $\bvec{t}$ is likewise one-dimensional. 
Its values can therefore be treated as constants in the eigenvalue computation, so that the characteristic function can be written directly as a function of $t$ as
\begin{equation}
\label{eq:charfct_1d}
    \varphi (t) = \prod_j (1-2i t \lambda_j)^{-1/2},
\end{equation}
such that the $\lambda_j$ 
\begin{equation}
\label{eq:eigvals_1d}
    \lambda_j \in \lambda \left(\bvec{M}_q^{\xi^+} \bvec{\Sigma}\right)
\end{equation}
only need to be computed once per pseudo-$a_{\ell m}$ covariance matrix, i.e. cosmology, and data dimension.
The characteristic-function grid $\bvec{t}$ is constructed from the mean and a specified number of standard deviations of the likelihood for each correlation-function dimension (see Secs.~\ref{sec:mean} and \ref{sec:cov}).

The eigenvalues can be computed from a simplified setup, similar to the product $\bvec{M}_q^{\xi^+} \bvec{\Sigma}$ as described in the beginning of Sec.~\ref{sec:implementation}.  
As long as the order of the pseudo-$a_{\ell m}$ vectors is respected, the pseudo-$a_{\ell m}$ covariance $\bvec{\Sigma}$ and combination matrix $\bvec{M}_q^{\xi^+}$ can always be rearranged such that the non-trivial selected blocks of $\bvec{\Sigma}$ are in the first two rows of blocks. 
Without loss of generality, we therefore denote the redshift bins under consideration for a cross-correlation as $z_1$ and $z_2$.
We write the characteristic polynomial as
\begin{equation}
    \label{eq:char_poly}
\det \bigl( \bvec{M}_q^{\xi^+} \bvec{\Sigma} - \lambda \mathbb{1} \bigr)  =
\det \begin{pmatrix} \bvec{T} & \bvec{Q} \\[2pt] \bvec{R} & \bvec{S} \end{pmatrix}
\end{equation}
with
\begin{equation}
\begin{aligned}
\bvec{T} &= \begin{pmatrix}
\bvec{m}_{\theta}\bvec{B}_{z_1,z_1}-\lambda\mathbb{1} & \bvec{m}_{\theta}\bvec{B}_{z_1,z_2} \\
\bvec{m}_{\theta}\bvec{B}_{z_2,z_1} & \bvec{m}_{\theta}\bvec{B}_{z_2,z_2}-\lambda\mathbb{1}
\end{pmatrix}, \\[6pt]
\bvec{Q} &= \begin{pmatrix}
\bvec{m}_{\theta}\bvec{B}_{z_1,z_3} & \cdots & \bvec{m}_{\theta}\bvec{B}_{z_1,z_n}\\
\bvec{m}_{\theta}\bvec{B}_{z_2,z_3} & \cdots & \bvec{m}_{\theta}\bvec{B}_{z_2,z_n}
\end{pmatrix}, \\[6pt]
\bvec{R} &= \begin{pmatrix}\mathbf 0 \end{pmatrix}, \\
\bvec{S} &= \mathrm{diag}(-\lambda\mathbb{1},\ldots,-\lambda\mathbb{1}),
\end{aligned}
\end{equation}
where we wrote out the row-wise multiplications of the combination vector $\bvec{m}_{\theta}$ with the sub-blocks $\bvec{B}_{z_i,z_j}$.
In terms of the matrices $\bvec{T}$, $\bvec{Q}$, $\bvec{R}$ and $\bvec{S}$, the determinant can be factorized as
\begin{equation}
    \det \bigl( \bvec{M}_q^{\xi^+} \bvec{\Sigma} - \lambda \mathbb{1} \bigr)
    = \det \bvec{S} \;\cdot\; \det \bigl( \bvec{T} - \bvec{Q} \bvec{S}^{-1} \bvec{R} \bigr)
    = (-\lambda)^{\,N_z} \, \det \bvec{T}. 
\end{equation}
The roots of this characteristic polynomial yield $N_z$ zero eigenvalues corresponding to the size of the sub-matrix $\bvec{S}$; these zeros do not contribute to the characteristic function of the likelihood, \cref{eq:charfct_1d}.
Only the eigenvalues of the submatrix $\bvec{T}$ affect the characteristic function. 
Consequently, one only needs to compute the eigenvalues of the products involving the pseudo-$a_{\ell m}$ covariance matrices of the two relevant redshift bins for each marginal.
For auto-correlations, the selection matrix picks out a single row/column, and the characteristic polynomial reduces to that of the corresponding pseudo-$a_{\ell m}$ auto-covariance for the redshift bin under consideration multiplied with the corresponding combination matrix.

\subsection{Covariance}
Following \cref{eq:xicov}, the covariance matrix for a set of correlation functions can be written as
\begin{equation}
    \operatorname{tr} \left( \bvec{M}^{\xi^+}_{m} \bvec{\Sigma} \bvec{M}^{\xi^+}_{n} \bvec{\Sigma}\right) = \sum \left( \bvec{M}^{\xi^+}_{m} \bvec{\Sigma} \circ \left(\bvec{M}^{\xi^+}_{n} \bvec{\Sigma}\right)^{\top}\right),
\end{equation}
where $\bvec{A} \circ \bvec{B}$ denotes element-wise (Hadamard) multiplication.
The sum is taken over all elements of this Hadamard product.

Multiplying a cross- or auto-correlation combination matrix with the pseudo-$a_{\ell m}$ covariance matrix results in one (auto) or two (cross) rows of selected covariance blocks multiplied with the combination matrix entries.
The blocks that contain only zeros in the first or second factor of the Hadamard product can be discarded.
We therefore obtain a tractable and compact way of writing the correlation function covariance matrix $\bvec{\Sigma}_{\xi}$ including all effects that are included in the pseudo-$a_{\ell m}$ covariance matrix $\bvec{\Sigma}$. 
This means in particular the mask shape and weights. 
\begin{multline}
\label{eq:xicov_det}
    \operatorname{cov} \left(\xi^+_{ij}(\bar{\theta}_1),\xi^+_{kl}(\bar{\theta}_2) \right) = 2 \operatorname{tr} \left( \bvec{M}_{ij}^{\theta_1} \bvec{\Sigma} \bvec{M}_{kl}^{\theta_2} \bvec{\Sigma}\right) \\ = \sum \left(\bvec{m}_{\theta_1} \bvec{B}_{ik} \circ \left(\bvec{m}_{\theta_2} \bvec{B}_{jl}\right)^{\top} + \bvec{m}_{\theta_1} \bvec{B}_{jk} \circ \left(\bvec{m}_{\theta_2} \bvec{B}_{il}\right)^{\top}\right),
\end{multline} 
where we went back to labelling redshift bins as $i$, $j$,... and angular bins explicitly for clarity.

We need to distinguish auto- and cross-correlations because the combination matrix will have twice the sidelength for the cross-correlations in order to combine auto- as well as cross-correlation pseudo-$a_{\ell m}$ covariance matrices.
In practice, we prepare all possible matrix products $\bvec{m}_{\theta_i} \bvec{B}_{jk}$ and combine them as needed to yield the covariance matrix entries for each data dimension. 
\\ 

\subsection{Approximations}
For computational feasibility, some approximations to the exact likelihood are necessary, which we will summarize in this section.
While the copula in itself already provides an approximation to the dependence structure of the full likelihood, other approximations enter through the choice and computation of the marginals as well as the covariance matrix. 
The approximations concerning the computation of the marginals and the covariance were already introduced in \citetalias{oehl2025a} and are described and justified there in detail.

\subsubsection{Non-Gaussian Marginals}
\label{sec:app_nongaussianmarginals}
Computing the pseudo-$a_{\ell m}$ covariance matrix $\bvec{\Sigma}$ as well as retrieving the eigenvalues from the products with the combination matrix is computationally expensive for high multipoles $\ell$.
This is because the sidelength of the covariance matrix scales as $\ell^2$.
We therefore approximate the high-$\ell$ part of the sum in the correlation-function estimator, \cref{eq:xip_alm}, to have a Gaussian sampling distribution, which is well motivated due to the central limit theorem.
In \citetalias{oehl2025a}, appendix G, we demonstrate which thresholds $\ell_{\mathrm{exact}}$ for this split are permissible while still representing the non-Gaussian sampling distributions well.
We then multiply the exact, low-$\ell$ characteristic function with the characteristic function of a Gaussian distribution for the high-$\ell$ part corresponding to a convolution in real (PDF) space.
In the following, we will nevertheless refer to these marginals as `exact marginals’ to distinguish them from fully Gaussian marginals.

\subsubsection{Covariance}
The same split in $\ell$ is applied to the correlation function covariance matrix $\bvec{\Sigma}_{\xi}$, where the low-$\ell$ part is computed exactly as in \cref{eq:xicov_det} and the high-$\ell$ part is computed in the $f_{\mathrm{sky}}$ approximation using the full-sky power spectra.
The full covariance is then given as the sum of the two parts.
Strictly speaking, the covariance can only be expressed as such a simple sum if the two parts are independent. 
However, computing the covariance between all $\ell$ is computationally intractable, which is why we adopt this approximation.
Instead, we verify that the multivariate PDFs constructed with the copula and using this approximation for the covariance represent the sampling distributions of a set of simulations well enough.

\subsubsection{Combining Different Marginal Shapes}
Within the copula approach it is possible to choose different marginals for different data dimensions freely. 
As the Gaussian likelihood approximation is an excellent approximation on small scales \citep[see e.g.][]{joachimi2021a} due to the central limit theorem, we will introduce purely Gaussian marginals for the smaller, sub-degree angular scales. 
To determine the angular bins in the data vector for which marginals can safely be assumed to be Gaussian, we start from a copula with only non-Gaussian marginals and replace them by Gaussian ones with the same variance starting from the smallest angular scales.
We record the likelihood value at the mean likelihood data vector for each set of marginals to see which replacement changes the likelihood noticeably.
Where likelihood evaluations are stable when changing from exact to Gaussian marginals, the Gaussian approximation suffices.
We find that this threshold lies around \SI{20}{\arcmin} for a \SI{10000}{\sqd} mask, as shown in \cref{appfig:gaussthreshold}.

\subsection{Different Couplings}\label{sec:couplings}
Additionally to the Gaussian copula we will be employing for this analysis we also tested a Student-t coupling due to its potential to model stronger tail dependence. 
However, we find the Student-t coupling is numerically unstable for the weak-lensing data vectors.
We leave the exploration of different couplings to future work but note that the fit with a Gaussian coupling is already quite remarkable.
\\
\\

\section{Application to Weak Lensing Surveys}
\label{sec:application}
In the remainder of this work we will explicitly apply our copula likelihood framework to weak-lensing correlation functions. 
Specifically, we will consider two survey geometries: stage III surveys such as KiDS-1000 (\SI{1000}{\sqd}) and stage IV surveys such as LSST (\SI{10000}{\sqd}), each approximated as a circular mask.
Unless stated otherwise, the map resolution adapted for the masks and simulations is $n_{\mathrm{side}} = 256$.
The effects of more realistic survey geometries were explored in \citetalias{oehl2025a}.
For redshift and angular-separation bins, we apply a KiDS-1000-like \citep{asgari2021a} setup based on logarithmically spaced angular bins extending up to \SI{5}{\degree}.
As the non-Gaussian likelihood is primarily driven by the survey geometry, the precise choice of redshift binning is not expected to affect our conclusions.
We modify the angular-bin setup in two ways.
First, angular bins with lower edges below \SI{15}{\arcmin} are omitted in the fiducial configuration due to the resolution of the simulated maps, leading to an effective scale cut of \SI{0.45}{\degree}. In subsequent sections we additionally include smaller angular scales again using higher-resolution masks; however, non-Gaussianity is not expected to be relevant on these scales.
Second, we add an additional large-scale bin spanning \SI{5}{\degree} to \SI{11}{\degree} to specifically probe non-Gaussian effects on the largest angular scales.
This leaves us with the angular bins $\bar\theta_1 = [\ang{0.45},\ang{1.01}]$, $\bar\theta_2 = [\ang{1.01},\ang{2.25}]$, $\bar\theta_3 = [\ang{2.25},\ang{5.00}]$ and $\bar\theta_4 = [\ang{5.00},\ang{11.12}]$.

\subsection{Comparison to Simulations}\label{sec:sims_comp}
In this section, we compare the copula likelihood to a set of simulations.
Our derived likelihood should match the sampling distribution of these simulations as they are created with the same statistical assumptions.
Namely, we simulate correlated Gaussian random weak-lensing fields with \texttt{GLASS}\footnote{\url{https://github.com/glass-dev/glass}} \citep{tessore2023a} using power spectra predicted for a flat cold-dark-matter model with a cosmological constant ($\Lambda$CDM) from the Core Cosmology Library\footnote{\url{https://github.com/LSSTDESC/CCL}} \citep[CCL;][]{chisari2019a} using the KiDS-1000 redshift bins \citep{hildebrandt2021a}. 
For demonstration purposes, we only used the $\SI{10000}{\sqd}$ mask and did not add shape noise as the non-Gaussianity becomes more apparent without noise.
Correlation functions were measured on these masked fields according to the estimator \cref{eq:xip_alm}, where pseudo-$C_{\ell}$ were estimated using \texttt{HEALPix}\footnote{\url{http://healpix.sf.net}} \citep{gorski2005a,zonca2019a}.
\begin{figure*}
    \includegraphics[width=\linewidth]{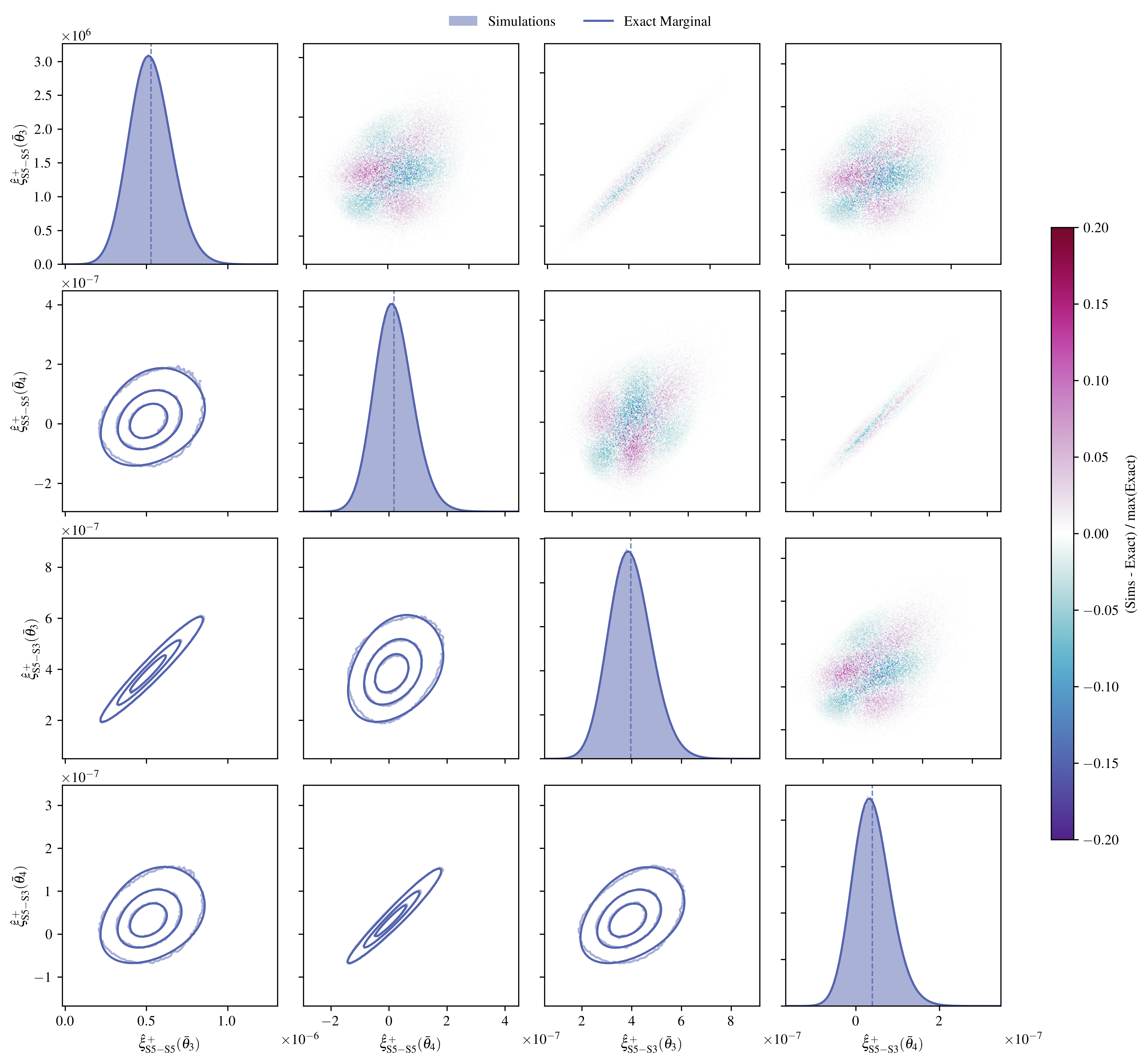}
    \vspace{-4mm}
    \caption{Comparison of two- and one-dimensional marginals of a set of $10^6$ simulations against the analytically predicted copula likelihood. The angular separations correspond to $\bar{\theta}_3 = [\ang{2.25},\ang{5.00}]$ and $\bar{\theta}_4 = [\ang{5.00},\ang{11.12}]$ and the redshift bins $S3$ and $S5$ are the same as in the KiDS-1000 analysis. Relative differences are shown in the upper triangle with respect to the maximum value of the copula PDF.
    }
    \label{fig:simscomp}
\end{figure*}
We selected two angular bins and two redshift-bin combinations and show the comparison of the one- and two-dimensional histograms of $10^6$ simulations to the analytically computed likelihood in \cref{fig:simscomp}.
The results look very similar for all other angular-separation bins and redshift-bin-combinations.
In the lower triangle we plot contours (enclosing $32 \%$, $68 \%$, $95 \%$) of the two-dimensional marginals of the full copula likelihood as solid lines.
The fainter lines show the same contours of the simulation histograms. 
We find that the copula likelihood captures the sampling distributions well. Applying a bootstrap to the histogram bin heights to quantify sampling noise due to the finite sample size, we find an average difference between the histogram bin heights and the copula likelihood evaluated at the bin centers of well below two sampling-noise standard deviations.

In the comparison panel (top right corner) of \cref{fig:simscomp}, deviations are shown as fractions of the maximum PDF value.
Especially in the highest-probability regions the agreement is excellent, such that the contours in the lower left corner of \cref{fig:simscomp} are barely distinguishable. 
The asymmetric shape of these two-dimensional contours highlights a clearly non-Gaussian structure that a fully Gaussian likelihood cannot capture. 
This is notable given that the copula assumes Gaussian coupling between dimensions, suggesting that the Gaussian coupling captures the dominant features of the dependence structure in this case.
For comparison, we show the same plot with a Gaussian likelihood in App.~\ref{app:simscomp_gauss}, where the deviations between sampling distribution and PDF are systematically larger and cannot be explained by sampling noise alone.
The number of outliers in bootstrap standard deviations per histogram bin is roughly twice as large as for the copula likelihood.

In \citetalias{oehl2025a} we showed that the non-linear scales contribute negligibly to the degree-scale correlation functions we are concerned with. 
We verify this here with an explicit comparison of correlation-function simulations produced with Gaussian and lognormal maps.
The lognormal simulations are run using GLASS with the same redshift-bin setup as for the Gaussian maps with a shift parameter as established in \citet{hilbert2011a}.
Tests confirm that the pseudo-$C_{\ell}$ measured on Gaussian and lognormal fields have the same mean.
The comparison of correlation-function simulations on the same angular scales as before using Gaussian and lognormal fields can be found in App.~\ref{app:simscomp_lognormal}.
The distributions are indistinguishable, which confirms the validity of the theoretically motivated assumption of sufficient Gaussianity of the weak-lensing fields at large enough scales.

\subsection{Posteriors}
In this section, we apply our exact likelihood to a realistic weak-lensing data vector to assess its impact on the resulting posteriors relative to a standard Gaussian likelihood.
We use the same setup as in the previous section, i.e. the KiDS-1000 redshift and angular bins described in Sec.~\ref{sec:sims_comp}.
The analysis is limited to $\xi^+$ data vectors as the large scales where the non-Gaussianity is most pronounced do not contribute to the $\xi^-$ data vectors significantly \citepalias{oehl2025a}.
To avoid introducing additional scatter unrelated to the likelihood choice, we use the mean predicted correlation function of a fiducial cosmology as our data vector.
For the Gaussian likelihood comparison, we adopt a covariance fixed at the fiducial cosmology, while the exact likelihood retains the parameter dependence described in Sec.~\ref{sec:cov}.
Unless noted otherwise, we adopt flat priors.
This choice is unimportant for our purposes, as identical priors are applied to both the Gaussian and copula likelihoods and our aim is to isolate the effect of the choice of likelihood rather than perform parameter inference; for comparison between the likelihoods, the simplest prior choice suffices.

\subsubsection{Fiducial Cosmology}
\begin{table}
\centering
\caption{Fiducial cosmological parameters and priors. Priors for either $\omega$ or $\Omega$ are used depending on the parameter set being sampled. }
\label{tab:cosmo_params}
\begin{tabular}{ccc}
\toprule
Parameter & Fiducial value & Prior \\
\midrule
$\omegam$ & 0.31 & $[0.1,\,0.5]$ \\ \\[-2.2ex]
$S_8 = \sigma_8(\omegam / 0.3)^{0.5}$                 & 0.80  & $[0.1,\,1.3]$\\ \\[-2.2ex]
$h$                   & 0.7 & $[0.64,\,0.82]$ \\ \\[-2.2ex]
$\omega_{\mathrm{b}}$ & 0.021 & $[0.019,\,0.026]$ \\ \\[-2.2ex]
$\omega_{\mathrm{c}}$ & 0.12 & $[0.051,\, 0.255]$ \\ \\[-2.2ex]
$n_{\mathrm{s}}$      & 0.97 & $[0.84,\, 1.1]$ \\ \\[-2.2ex]
\midrule
$A_{\mathrm{IA}}$     & $0$  & $[-6,\, 6]$ \\ \\[-2.2ex]
\midrule
$\delta_z$            & $\bvec{0}$ & $\mathcal{N}(\bvec{z}_{\mathrm{mean}},\,\bvec{\Sigma}_z)$ \\ \\[-2.2ex]
\bottomrule
\end{tabular}
\end{table}

For all posteriors we will be using fiducial values and priors as listed in \cref{tab:cosmo_params}, adopted from \citet{asgari2021a}.
For mean and covariance of the redshift distribution shift parameter $\delta_z$ we refer to \citet{hildebrandt2021a} and adopted the covariance values in \citet[][their fig.~7]{joachimi2021a}.
Priors for either $\omega$ or $\Omega$ are used depending on which set of parameters is sampled. 
\\
\\

\subsubsection{One-Dimensional Parameter Space}
\label{sec:onedim}
We begin by presenting one-dimensional posteriors for $S_8$, primarily for comparison with our earlier results based on a one-dimensional data vector \citepalias{oehl2025a}. 
Posteriors are computed on an evenly spaced parameter grid of $S_8$.
All other cosmological parameters are fixed to their fiducial values.
In contrast to our previous work, the copula framework allows us to use the full data vector rather than a single data point.
To match the assumptions of our previous work, we excluded angular bins below \SI{15}{\arcmin}, used a \SI{10000}{\sqd} mask and no shape noise.
The high signal-to-noise ratio was necessary to produce a constraint with only one data point in our previous work.

\begin{figure}
    \centering
    \includegraphics[width=\linewidth]{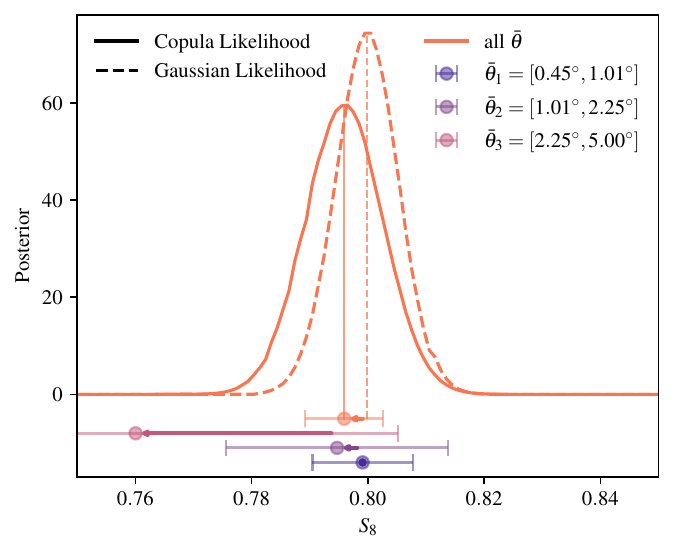}
    \caption{Posteriors comparing the use of a fully Gaussian (dashed line) to the copula likelihood (solid line) using a \SI{10000}{\sqd} mask and a mock KiDS-$\xi^+$-data vector excluding scales below \SI{15}{\arcmin}.
    In each case all redshift-bin combinations are included, while different colors indicate the use of different angular bins.
    The orange lines show the posteriors when all angular bins are included. 
    Single angular bins to compare to our previous work \citepalias{oehl2025a} are shown as horizontal lines indicating one standard deviation around the mean (circle) when using the copula likelihood. 
    The corresponding shifts of the means from the Gaussian likelihood means are indicated by the arrows.}
    \label{fig:post1d_10000}
\end{figure}

In \cref{fig:post1d_10000}, we show the one-dimensional posteriors on $S_8$ using the copula likelihood (solid line) and the Gaussian likelihood (dashed line). 
The orange lines show posteriors using all five redshift bins and all angular bins.

Additionally, we include means and standard deviations of posteriors inferred with all $15$ redshift-bin combinations but only one angular bin at a time as colored lines below the main plot. 
For comparison, the purple line ($\bar{\theta}_2$) corresponds to the angular bin that has been used in \citetalias{oehl2025a}.
This result shows the effect of adding correlations between redshift bins but not changing the angular scale.

We find that adding correlations between redshift bins changes the direction of the shift of the mean with respect to the Gaussian likelihood compared to our previous results using only one redshift-bin combination \citepalias{oehl2025a}.
In each case, the mean of the posterior obtained with the copula likelihood is shifted to lower values of $S_8$ compared to the Gaussian likelihood.
These shifts are illustrated with arrows below the posteriors and the transparent ranges correspond to the $68\%$ confidence intervals of each posterior obtained with the copula likelihood.
As expected, smaller angular bins yield stronger constraining power and the choice of the likelihood has less of an impact as small angular scales are more Gaussian distributed due to the central limit theorem.
In particular, we anticipated from this result that sub-degree marginals can safely be modelled as Gaussian but present more rigorous results on the replacement of exact marginals with Gaussian marginals in App.~\ref{app:gauss_margins}.

Using all angular bins (orange lines) further increases the constraining power but increases the shift in the posterior mean again due to the inclusion of the large non-Gaussian scales.
Notably, the shift is now on the order of magnitude of one standard deviation of the posterior.

\subsubsection{Two-Dimensional Parameter Space}
\label{sec:twodim}
In this section, we present posteriors in a two-dimensional parameter space, namely $\omegam$ and $S_8$.
All other parameters are fixed to their fiducial values.
As in the one-dimensional results on $S_8$, we compute the two-dimensional posteriors by evaluating the likelihood on an evenly spaced grid in parameter space.
Priors are adopted as noted in \cref{tab:cosmo_params}, where the ranges of the $S_8$ priors are adapted to resolve each posterior well.
Throughout this section, we use only exact marginals, i.e. none are replaced by fully Gaussian marginals, as the computational cost is manageable in this setup. 
For computational feasibility, the exact marginals themselves are computed using an exact treatment up to $\ell_{\mathrm{exact}} = 20$, with higher multipoles incorporated through the convolution described in Sec.~\ref{sec:app_nongaussianmarginals}. This threshold is on the low side for a \SI{1000}{\sqd} mask but sufficient for a \SI{10000}{\sqd} mask \citepalias[see][]{oehl2025a}.
This means that differences in the posteriors could be exacerbated by taking the $\ell_{\mathrm{exact}}$ threshold to a higher multipole moment.

We also include shape noise here, since we are no longer comparing to the previous one-dimensional results but instead aim to assess the impact of the likelihood choice on realistic posteriors.
Specifically, we add shape noise corresponding to a galaxy number density of $n_{\mathrm{gal}} = \SI{1.21}{\per \arcmin \squared}$ per redshift bin 
and an ellipticity dispersion of $\sigma_{\epsilon,1/2} = 0.28$ per component, consistent with KiDS-1000 \citep{giblin2021a} and broadly comparable to expected LSST Year~1 data for a single redshift bin.

\begin{figure*}
    \centering
    \includegraphics[width=\linewidth]{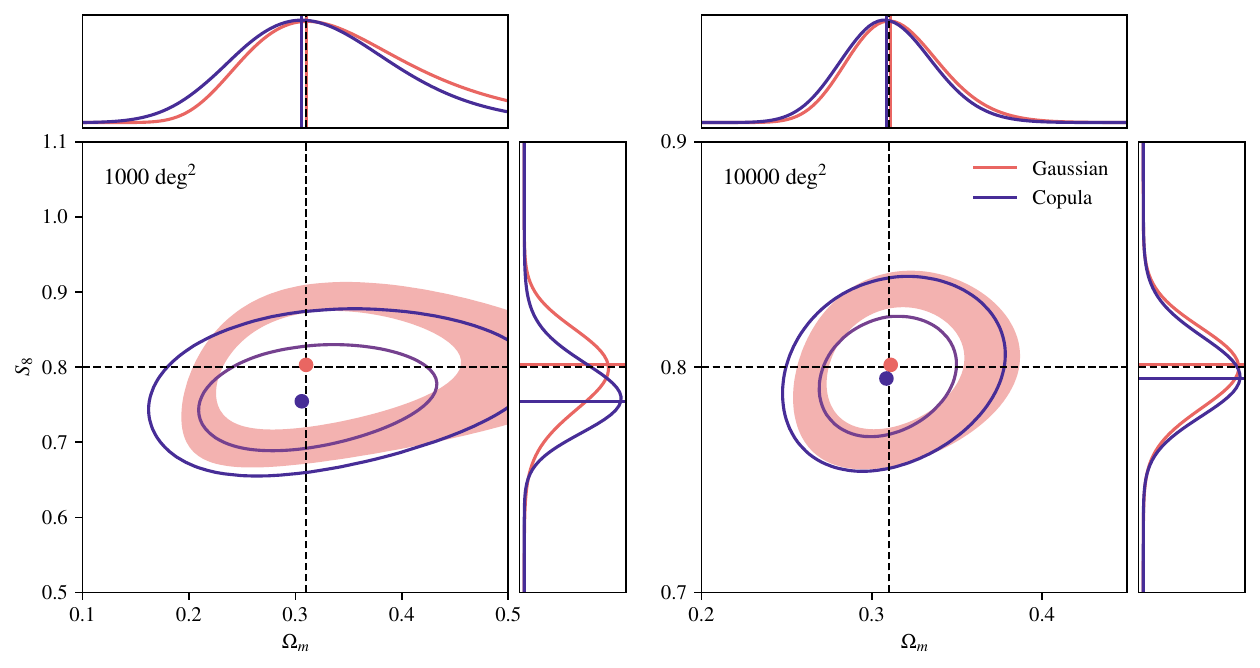}
    \caption{Two-dimensional posteriors in the $\omegam$-$S_8$ plane using a \SI{1000}{\sqd} mask (left panel) and a \SI{10000}{\sqd} mask (right panel). For both posterios, a full simulated KiDS-$\xi^+$-data vector excluding small scales $< \SI{30}{\arcmin}$ has been used and the $68\%$ and $95\%$ confidence levels are shown as contours.
    Black dashed lines mark the fiducial cosmology, while dots or solid lines indicate the modes.}
    \label{fig:post2d_maskcomp}
\end{figure*}

\begin{figure*}
    \centering
    \includegraphics[width=\linewidth]{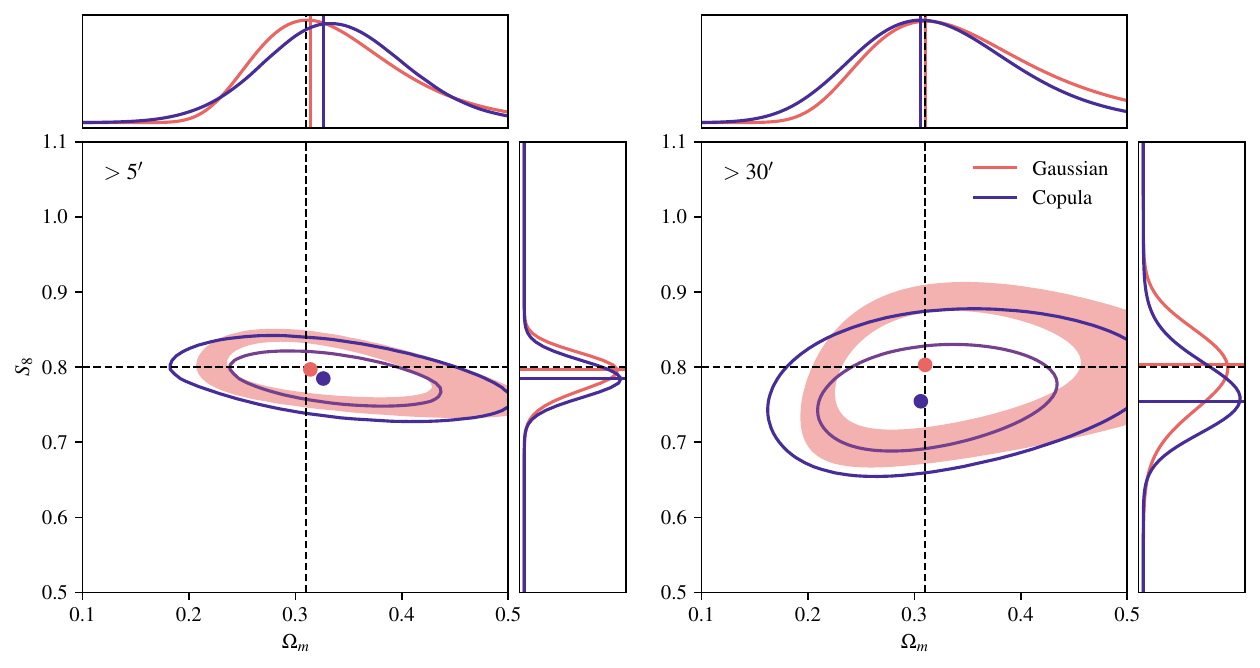}
    \caption{Two-dimensional posteriors in the $\omegam$-$S_8$ plane using a \SI{1000}{\sqd} mask. A simulated KiDS-$\xi^+$-data vector is used for both panels, including angular bins down to \SI{5}{\arcmin} (left panel) and down to \SI{30}{\arcmin} (right panel).
    Contours show the $68\%$ and $95\%$ confidence levels and the fiducial cosmology is indicated by the black dashed lines. 
    Note that the right panel corresponds to the left panel in \cref{fig:post2d_maskcomp}.}
    \label{fig:post2d_scalecomp}
\end{figure*}
We illustrate our results through two comparisons:
(i) two survey masks while using the same data vector, and (ii) two data vectors (with and without small-scale information) evaluated with the same survey mask.
Complementary cases are shown in App.~\ref{app:comps}.

We show the results of case (i) in \cref{fig:post2d_maskcomp}.
Here, we used a simulated KiDS-$\xi^+$-data vector excluding scales below \SI{30}{\arcmin} and compare the \SI{1000}{\sqd} mask (left panel) to the \SI{10000}{\sqd} case (right panel).
Posteriors obtained with the copula likelihood are shown as purple contours ($68\%$ and $95\%$ confidence levels).
For comparison, posteriors obtained with the corresponding Gaussian likelihood using a constant covariance matrix are shown as pink contours. 
Modes are indicated in the same colors as lines and dots in the marginals and two-dimensional contours respectively.
Additionally, black dashed lines are drawn at the fiducial values.
As in the one-dimensional posteriors shown in Sec.~\ref{sec:onedim}, the mean and the mode of the $S_8$ posterior sit slightly lower using the exact likelihood for the \SI{1000}{\sqd} mask.
In the $\omegam$-direction the mode is not shifted but the mean as derived with the copula likelihood is slightly higher.
The two posteriors clearly differ in shape, and the shift of the $S_8$ mode is of order one standard deviation.

In this analysis we use the expectation value of the estimator as the data vector, corresponding to the mean of many realizations.
For the skewed sampling distribution described by the copula likelihood, this mean does not coincide with the most likely realization. 
Consequently, the fiducial cosmology is not expected to be recovered at the mode of the posterior when using this particular data vector. 
For a Gaussian likelihood, however, mean and mode coincide, such that the expectation value represents the most likely data vector for a given cosmology and the fiducial cosmology is therefore recovered at the mode of the posterior. 
For the larger survey area of \SI{10000}{\sqd}, the difference between the two likelihoods becomes negligible even when excluding small scales.

In comparison case (ii), \cref{fig:post2d_scalecomp}, we used the \SI{1000}{\sqd} mask and added smaller angular bins down to \SI{5}{\arcmin} (left panel).
We use this to demonstrate the effect of the number of data points and angular scales used on the posteriors. 
The copula likelihood (purple contours) is again compared to the Gaussian likelihood (pink contours). 
The posteriors of the right panel correspond to the same physical setup as in \cref{fig:post2d_maskcomp}, left panel, but on a technical level, the exact marginals and covariances are computed using a higher mask resolution ($n_{\mathrm{side}}=1024$) for \cref{fig:post2d_scalecomp} to resolve the smaller angular bins.
Even though strictly not necessary for the right panel, this higher resolution is kept to be consistent with the left panel and as an additional check.

We refrain from going to even smaller scales as the computational cost increases while the additional marginals are far in the Gaussian regime \citep{joachimi2021a} and are therefore not expected to add any changes when comparing to a standard Gaussian likelihood.
Additionally, on even smaller scales, baryonic feedback affects the signal, such that those scales are often excluded from weak lensing analyses \citep[for example][]{dark-energy-survey-and-kilo-degree-survey-collaboration2023a}.

Adding the smaller angular-separation bins improves the constraining power as expected and reduces the difference in posteriors obtained from the Gaussian and copula likelihood.
The differences between the left and right panels of \cref{fig:post2d_scalecomp} are also interesting in light of \citet[][their fig.~2]{stolzner2025a}, who show slight shifts in the $S_8$–$\omegam$ posteriors when varying the angular-scale range.
When only using large scales ($>\SI{18.5}{\arcmin}$), their $S_8$ posterior is slightly lower than when using all scales or only small scales.
However, in our large-scale-only configuration, the copula likelihood shifts the posterior down compared to the Gaussian likelihood and not back up, meaning that the use of a Gaussian likelihood on these large scales cannot be the sole cause of the shift \citet{stolzner2025a} observe.

Both of these comparisons show that the impact of the exact likelihood on posteriors is highly sensitive to the particular choice of survey mask and scales included.

\subsubsection{Sampling of the Full Parameter Space}
\label{sec:sampling}
\begin{figure}
    \centering
    \includegraphics[width=\linewidth]{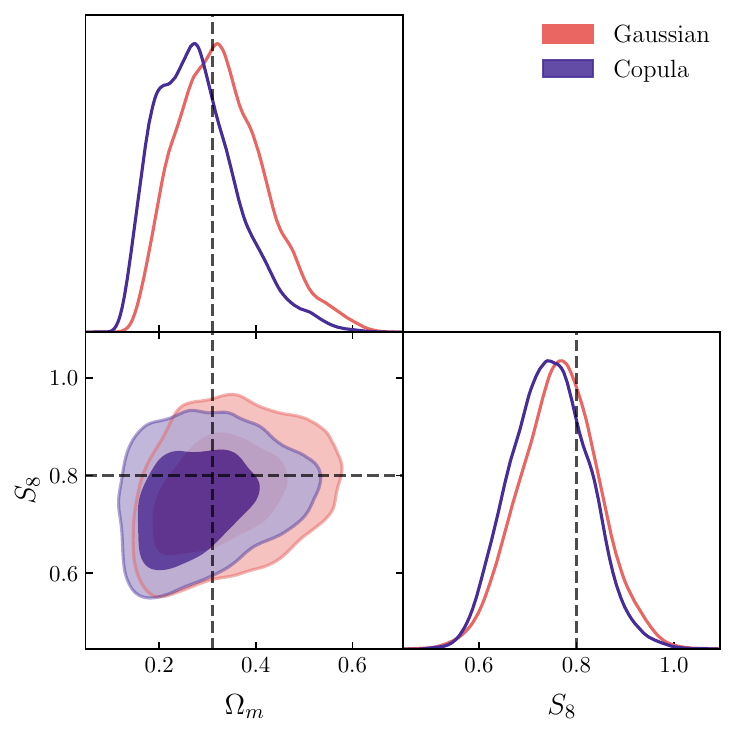}
    \caption{Two-dimensional posteriors obtained from sampling the Gaussian (pink contours) and copula likelihood (purple contours) in a full weak lensing parameter space. We show marginals in the $\omegam$-$S_8$ plane. A simulated KiDS-$\xi^+$-data vector including angular bins down to \SI{30}{\arcmin} and  a \SI{1000}{\sqd} mask were used.
    Contours show the $68\%$ and $95\%$ confidence levels and the fiducial cosmology is indicated by the black dashed lines.}
    \label{fig:sampled_posterior}
\end{figure}
Having explored the impact of the copula likelihood on the posteriors of the main cosmological parameters constrained by weak lensing on a fixed grid, we conclude with a proof-of-principle example showing that the framework can also be used when sampling the full parameter space, thus demonstrating the applicability of the framework to full-fledged analyses. 
We sample the posterior using \texttt{emcee} \citep{foreman-mackey2013a}.
We adopt the fiducial stage-III-like setup introduced at the beginning of Sec.~\ref{sec:application}, using the \SI{1000}{\sqd} mask with $n_{\mathrm{side}} = 1024$. Relative to the original KiDS-1000 analysis setup, this configuration emphasizes larger angular scales where non-Gaussian likelihood effects are expected to be most pronounced.
We vary the same cosmological and nuisance parameters as \citet{asgari2021a}, except for the baryonic feedback nuisance parameter and additive ellipticity bias, and adopt the same priors listed in \cref{tab:cosmo_params}.
We show the main results in \cref{fig:sampled_posterior}. 
Comparing posteriors obtained with a Gaussian and our copula likelihood, the shifts are qualitatively similar to what was observed in \cref{fig:post2d_maskcomp}.
This was expected since the mock data and survey setup are the same.
However, it is reassuring to see that shifts do not seem to be affected by the number of parameters varied. 
The corner plot showing all parameters can be found in App.~\ref{app:fullsampling}.

\section{Discussion \& Conclusion}
In this work, we explored exact non-Gaussian likelihoods for weak-lensing correlation functions.
We demonstrated how to construct an approximation to the correlation-function likelihood using a copula framework with exact non-Gaussian marginals.
The copula likelihood matches simulated sampling distributions of correlation functions well, with differences between the sampling distribution and the analytical joined PDF much smaller for the copula likelihood than the Gaussian likelihood.

We showed results for weak-lensing posteriors comparing the use of the copula likelihood and the standard Gaussian likelihood.
Using the more accurate copula likelihood induces shifts in the posterior constraints on $\omegam$ and $S_8$ due to its skewed nature. 
The magnitude of these shifts crucially depends on the geometry of the survey and the structure of the data vector, such as the number of data points and correlation between them.
For large survey masks, the differences between using a Gaussian and the copula likelihood are small for posteriors of cosmological parameters.
Based on our findings, we nevertheless recommend running a test on the $S_8$ inference using our framework with the final footprint and data vectors for stage IV weak-lensing surveys like LSST or Euclid, as the exact mask geometry and data vector structure, not just overall size, can make a difference.
Our framework can readily be applied to real weak-lensing data as it allows for a general continuous weighting of the fields, not just binary masks.

Our analyses are limited to a Gaussian coupling and simple masks. 
We leave rigorous testing and stable implementation of other couplings for future work.
In this work, we applied the copula likelihood to $\xi^+$-data vectors only as these are most influenced by the large scales, as has been shown in previous work \citepalias{oehl2025a}.
We leave adding $\xi^-$ for completeness to future applications of our copula framework.

The copula framework is general and can be used for any non-Gaussian likelihood.
This includes combined galaxy clustering and weak lensing analyses as well as any other two-point statistics.
Moreover, the framework will be suitable to other cosmological probes, such as two-point functions of \SI{21}{\centi \metre} maps, the Hellings-Downs curve for the stochastic gravitational-wave background and any other summary statistic with known one-dimensional marginals.

\section*{Acknowledgements}
\small
We acknowledge funding from the Swiss National Science Foundation under the Ambizione project PZ00P2\_193352.

\textbf{Software}: For this work, we made use of the Python packages \texttt{numpy} \citep{harris2020a}, \texttt{scipy} \citep{virtanen2020a}, \texttt{jax} \citep{jax2018github}, \texttt{Wigner}\footnote{https://github.com/ntessore/wigner} \citep[based on][]{schulten1975a}, \texttt{pyCCL}, \texttt{GLASS}, \texttt{healpy} and figures were created with \texttt{Matplotlib} \citep{hunter2007a} and \texttt{GetDist} \citep{Lewis:2019xzd}.

\bibliographystyle{aasjournal}
\bibliography{library}
\newpage

\appendix
\Crefname{section}{App.}{Apps.}

\section{Choice of the Threshold for Gaussian Marginals}
\label{app:gauss_margins}
The copula approach allows to employ Gaussian marginals next to the computationally expensive exact marginals wherever appropriate.
The non-Gaussianity of the likelihood is most prominent on large scales, so the smaller the angular separation bin, the better the Gaussian approximation becomes \citep[e.g.][]{joachimi2021a,hall2022a}.
To find the angular size below which Gaussian marginals can be safely employed, we successively replaced exact marginals going into the copula construction by Gaussian marginals, starting with the smallest scales. 
For these tests including smaller angular scales, higher-resolution masks with $n_{\mathrm{side}} = 1024$ are used to resolve the corresponding bins.
\begin{figure}
\centering
    \includegraphics[width=\linewidth]{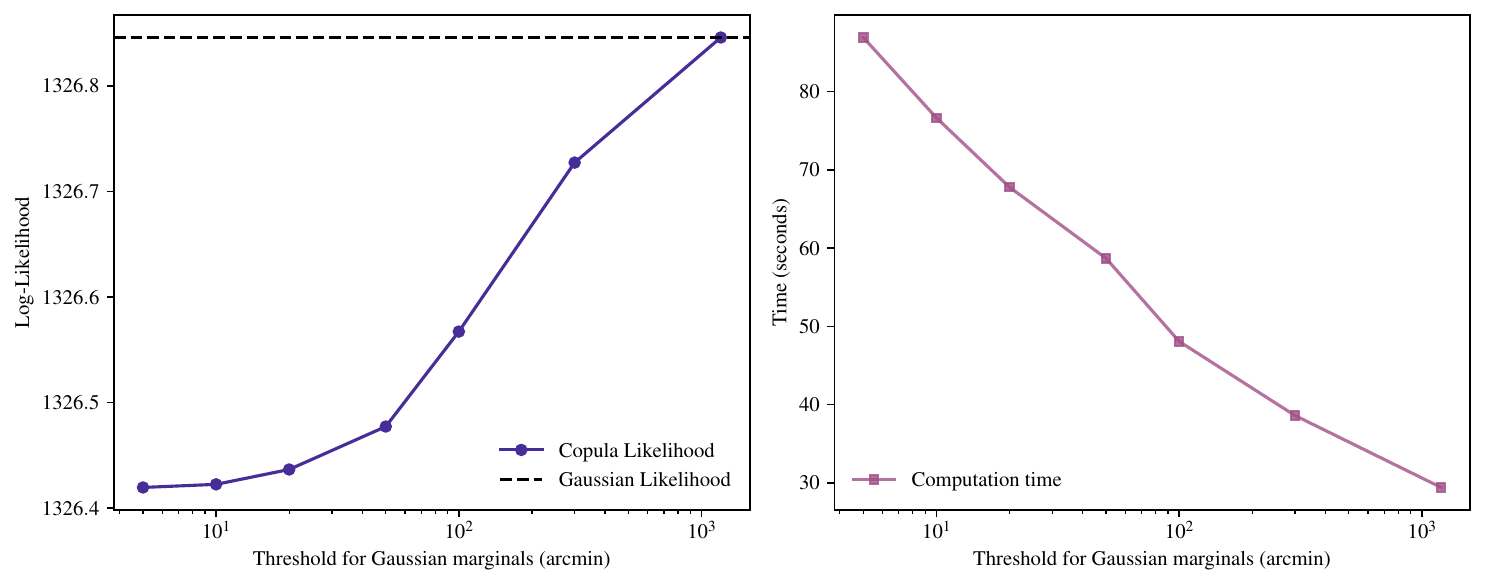}
    \vspace{-4mm}
    \caption{Marginals used in a copula can be of mixed shapes. We show successive replacement of small-angular-scale exact marginals by Gaussian marginals, where the threshold for the replacement is shown on the horizontal axis. In the left panel, the log-likelihood value is plotted, indicating a significant change from \SI{20}{\arcmin} onwards. In the right panel, the corresponding computation time for one likelihood evaluation is shown.}
    \phantomsection
    \label{appfig:gaussthreshold}
\end{figure}
We show the resulting likelihood values for our fiducial cosmology and data vector in \cref{appfig:gaussthreshold}.
The thresholds on the horizontal axis set the minimum lower angular-bin edge for a marginal to be computed exactly: all marginals with a lower angular-bin edge below this threshold are replaced by Gaussian marginals.
As we are only interested in any change in the likelihood, the data vector and set of cosmological parameters used for the evaluation are irrelevant, as long as kept the same for each configuration of exact and Gaussian marginals. 
The computational time to evaluate the likelihood decreases as more exact marginals are replaced by Gaussians as shown in the right panel of \cref{appfig:gaussthreshold}. 

\section{Comparison of Gaussian likelihood to Simulations}
\label{app:simscomp_gauss}
Our copula likelihood represents sampling distributions of weak-lensing correlation functions well (see \cref{fig:simscomp}).
\begin{figure*}
    \includegraphics[width=\linewidth]{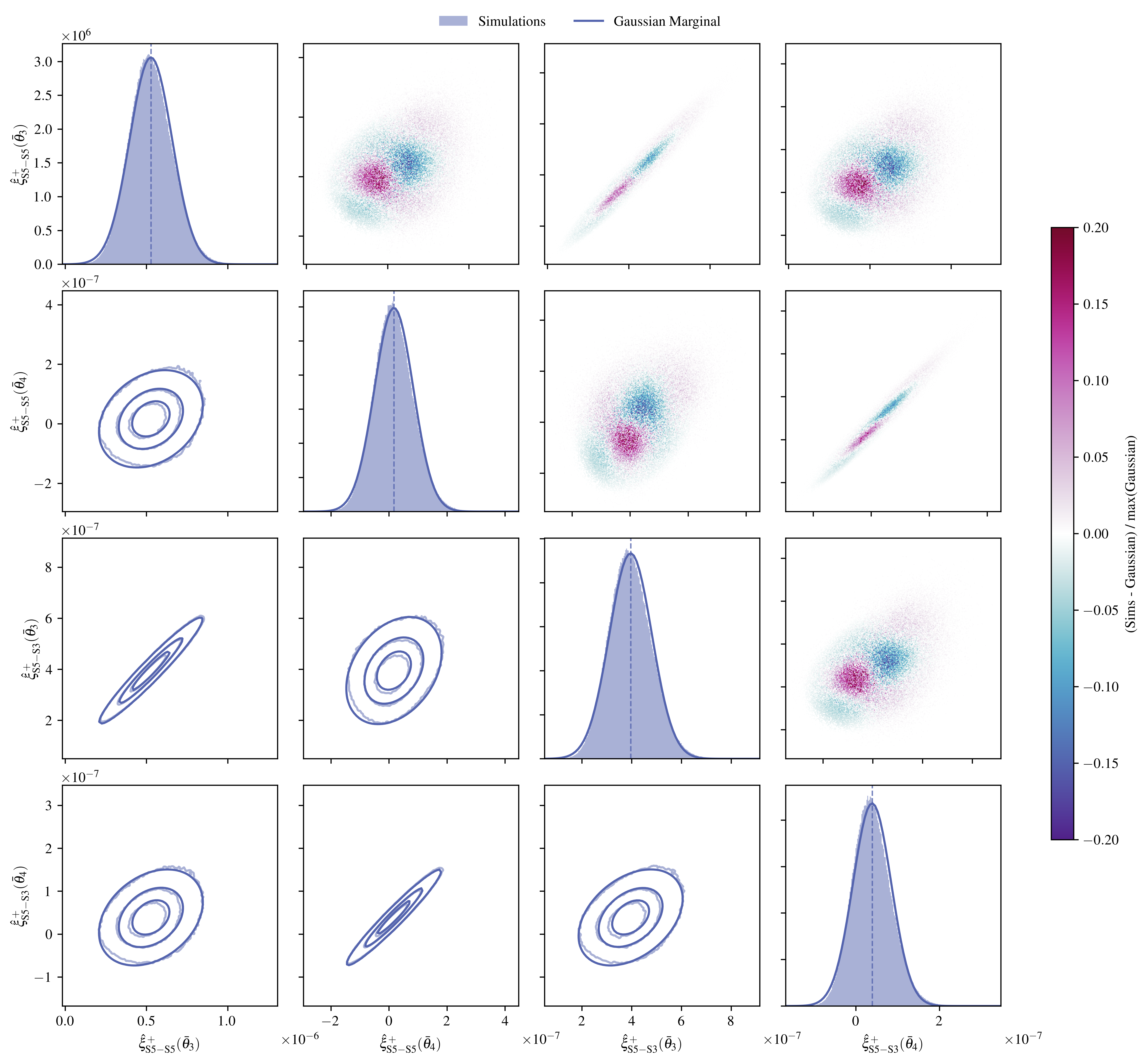}
    \vspace{-4mm}
    \caption{Comparison of two- and one-dimensional marginals of a set of $10^6$ simulations against the analytically predicted Gaussian likelihood. The angular separations correspond to $\bar{\theta}_3 = [\ang{2.25},\ang{5.00}]$ and $\bar{\theta}_4 = [\ang{5.00},\ang{11.12}]$ and the redshift bins $S3$ and $S5$ are the same as in the KiDS-1000 analysis. Relative differences are shown in the upper triangle with respect to the maximum value of the exact PDF. The one-dimensional marginals of the simulations show a systematic offset to the Gaussian PDF on the diagonal and the two-dimensional marginals in the lower left corner also show this slight offset, particularly in the high density regions.}
    \phantomsection
    \label{fig:simscomp_gauss}
\end{figure*}
In \cref{fig:simscomp_gauss} we show the same histograms of simulations but compare them to a multivariate Gaussian distribution instead of the copula with exact marginals. 
It is evident that the level of agreement between simulations and analytical PDF is worse than with the copula, i.e. differences in the upper right corner are more pronounced. 
In absolute terms, this can be seen on the diagonal and in the lower-left triangle, where contour lines are now more distinguishable. 
Differences are not only larger in absolute terms but also with respect to the sampling noise level determined using a bootstrap on the histograms as explained in Sec.~\ref{sec:sims_comp}.

\section{Comparison of Gaussian and Lognormal Simulations}\label{app:simscomp_lognormal}
We compared our correlation-function simulations obtained from Gaussian random fields to an ensemble of correlation functions obtained from lognormal fields.
\begin{figure*}
    \includegraphics[width=\linewidth]{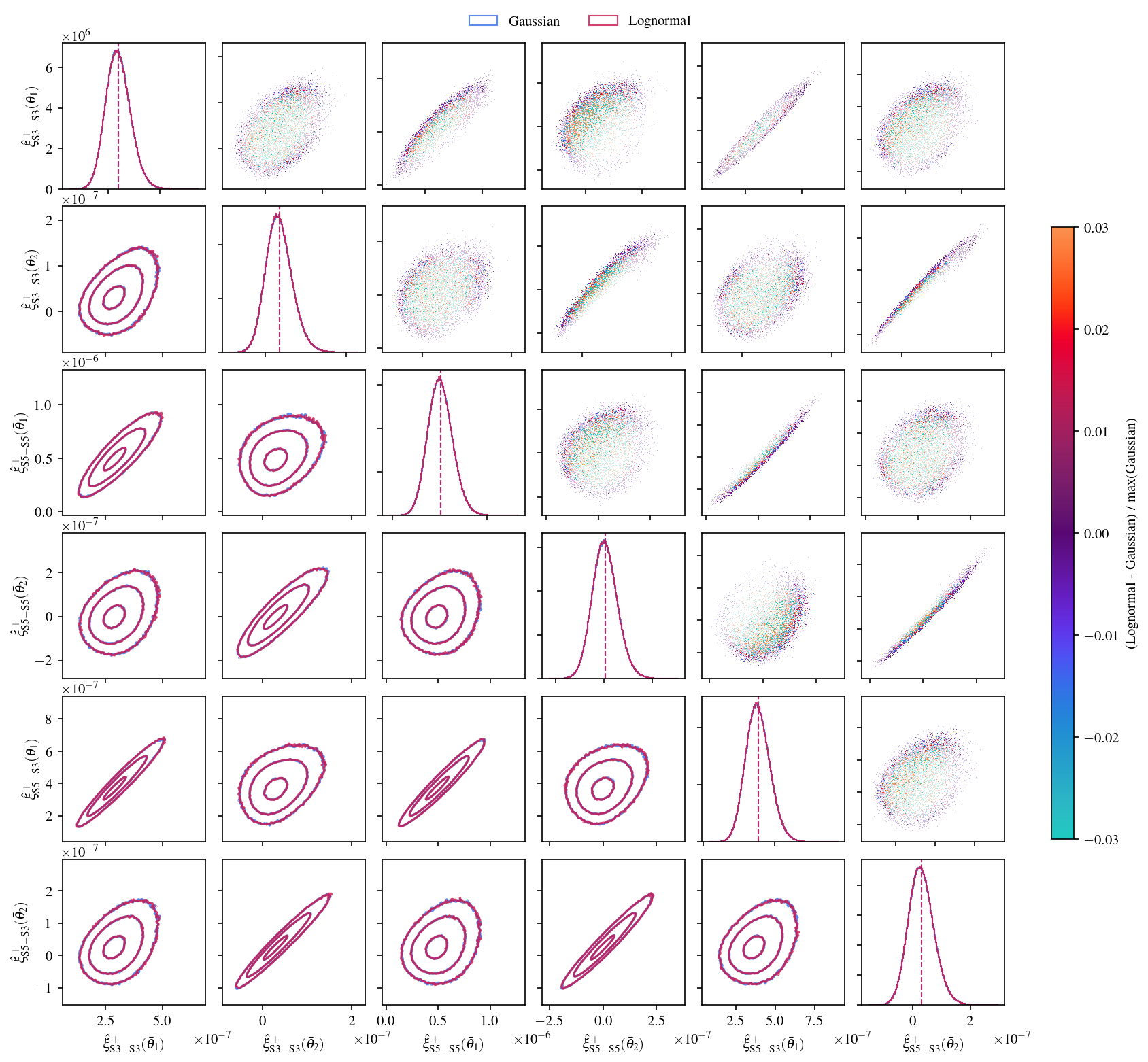}
    \vspace{-4mm}
    \caption{Comparison of two- and one-dimensional marginals of two sets of $10^6$ correlation-function simulations obtained from Gaussian and lognormal random fields. The angular separations correspond to $\bar\theta_1 = [\ang{0.45},\ang{1.01}]$ and $\bar\theta_2 = [\ang{1.01},\ang{2.25}]$, showing good agreement even on the smaller scales we use, and the redshift bins $S3$ and $S5$ are the same as in the KiDS-1000 analysis.}
    \phantomsection
    \label{fig:simscomp_lognormal}
\end{figure*}
These lognormal fields were generated with GLASS. 
The direct comparison in terms of one- and two-dimensional histograms can be seen in \cref{fig:simscomp_lognormal}. 
Opacity is an indicator for statistical significance in terms of standard deviations of boostrap resamplings of the histograms. 
If the difference between the bin heights of correlation functions computed on Gaussian and lognormal fields is smaller than one standard deviation, the bin will be marked completely transparent. 
Between one and three standard deviations it is slightly transparent and everything above three standard deviations fully opaque.
The colors show bin differences with respect to the maximum of correlation-function distributions measured on Gaussian random fields. 

\section{Additional comparisons}
\label{app:comps}
\begin{figure*}
    \includegraphics[width=\linewidth]{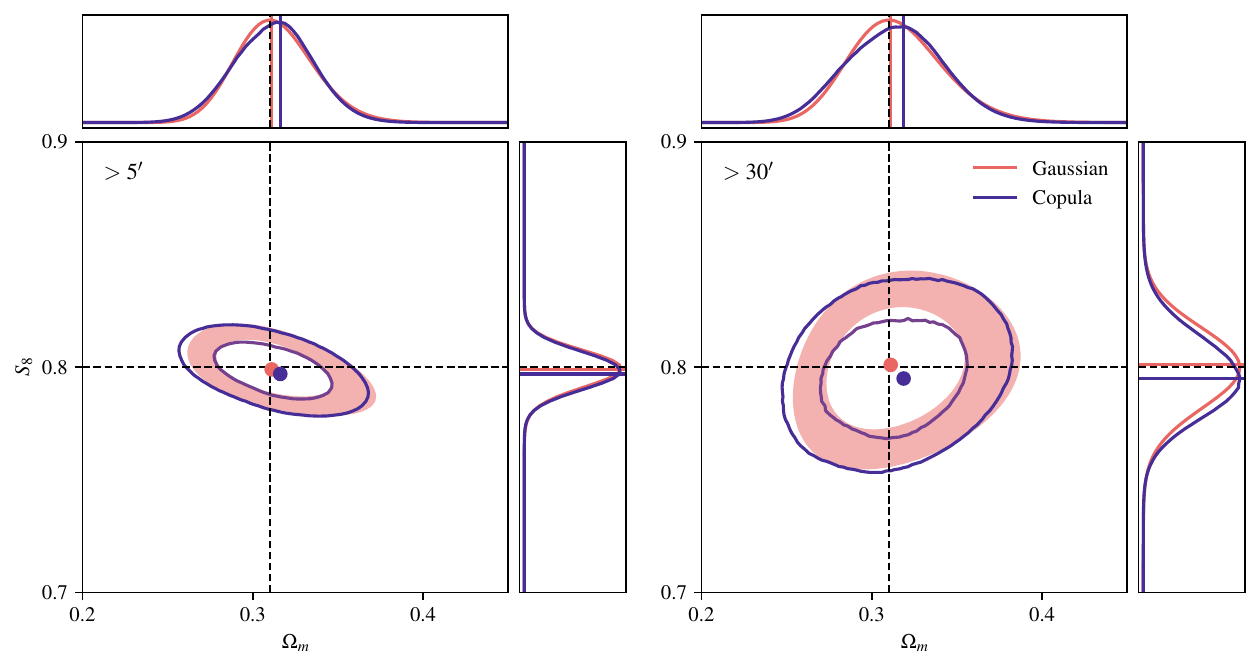}
    \vspace{-4mm}
    \caption{Two-dimensional posteriors comparing different scale-cuts as in \cref{fig:post2d_scalecomp} but employing a \SI{10000}{\sqd} mask.}
    \phantomsection
    \label{fig:2dpost_10000}
\end{figure*}

\Cref{fig:2dpost_10000} shows two-dimensional posteriors obtained from a different range of scales. The setup is the same as for \cref{fig:post2d_scalecomp} but now using a \SI{10000}{\sqd} mask. 
The differences in the posteriors are qualitatively the same as in \cref{fig:post2d_scalecomp} but almost negligible in magnitude. 

\section{Full parameter space posteriors}
\label{app:fullsampling}
\begin{figure*}
\centering
    \includegraphics[width=\linewidth]{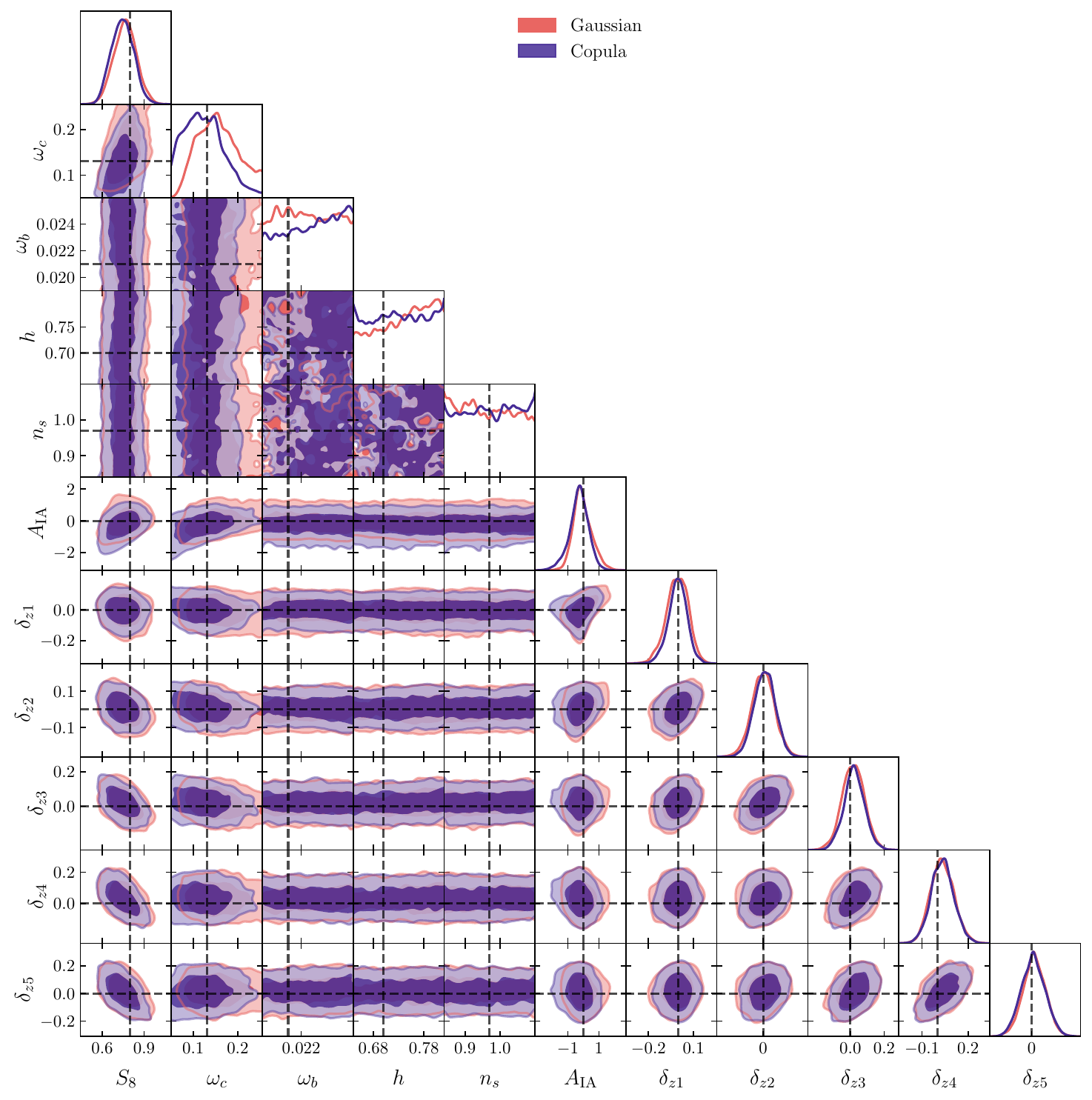}
    \vspace{-4mm}
    \caption{Posteriors for a full set of weak-lensing parameters including the cosmological parameters (left five) as well as the linear intrinsic alignment parameter $A_\mathrm{IA}$ and shift parameters for the five redshift bins $\delta_z$.}
    \phantomsection
    \label{fig:full_grid}
\end{figure*}
We show all parameters sampled in our sampling demonstration using the copula likelihood framework (Sec.~\ref{sec:sampling}) in \cref{fig:full_grid}.
For a detailed description of all parameters we refer to \citet{asgari2021a}.

\end{document}